# Polycrystalline ferroelectric croconic acid for multisource environmental energy harvesting

*Gloria P. Moreno-Martínez, Hari K. Mishra,* Xabier García-Casas, María Alcaire, Vanda Godinho, Juan R. Sánchez-Valencia, Ana Borrás, Angel Barranco**

Nanotechnology on Surfaces and Plasma Group, Materials Science Institute of Seville (CSIC-US), c/Américo Vespucio 49, 41092 Seville, Spain

***e-mail:** hari.krishna@icmse.csic.es; angel.barranco@csic.es

## Abstract

The development of organic ferroelectric materials with scalable, simplified fabrication routes remains a key challenge for next-generation energy-harvesting technologies. In this context, croconic acid emerges as a distinctive candidate, exhibiting a proton transfer-driven polarization mechanism that enables large spontaneous polarization through collective hydrogen-bond rearrangements rather than conventional ionic displacement. In this work, polycrystalline croconic acid (CA) thin films are fabricated by vacuum sublimation on Ar plasma-treated flexible substrates and then stabilized via in situ encapsulation with an adamantane-based remote plasma polymer. This integrated approach effectively suppresses surface degradation under ambient conditions, enabling long-term stability for more than one year. The films exhibit robust ferroelectric behavior with an oblique polarisation orientation, as confirmed by piezoresponse force microscopy, revealing well-defined domains and localized low coercive fields at the nanoscale. The functional electrical response of these films is demonstrated by integrating them into multilayer piezoelectric and pyroelectric systems in response to mechanical and thermal stimuli, respectively. Notably, the piezoelectric response is strongly influenced by film thickness, leading to enhanced device performance for thicker films. Besides, embedding the CA between dielectric polymeric layers enhances macroscopic device performance, reaching maximum power density outputs as high as 37 μW/m$^2$ for ~2 μm of piezoelectric counterpart. Crucially, despite the convention that high structural crystallinity is imperative for sustaining ferro-, piezo-, and pyroelectricity, our polycrystalline CA thin films demonstrate an exceptional room-temperature pyroelectricity, which has not been explored in CA-based devices to date. The CA thin film exhibits a pyroelectric coefficient of ~10 $\mu C/m^2.K$, which underscores a robust functional response that rivals established benchmarks in both organic and inorganic pyroelectric material classes. Thus, CA organic ferroelectric thin films emerge as viable candidates for efficient thermal energy harvesting. Beyond the functional properties, the proposed solvent-free, low-temperature fabrication route, which combines deposition and encapsulation in a single in situ process, represents a significant step toward simplifying the integration of organic lead-free ferroelectric materials into practical devices. Its compatibility with scalable vacuum technologies and flexible substrates, along with ample room for optimization, opens the way to low-cost, high-performance multisource energy harvesting systems for the next generation.

## 1. Introduction

Developing self-powered smart systems requires a shift away from single source energy transducers toward versatile materials capable of multisource environmental energy harvesting.[1,2] Multisource energy harvesting, specifically the simultaneous conversion of mechanical vibrations and ambient thermal fluctuations, offers a robust pathway towards continuous power generation by maximizing energy density within a single device footprint.[3] Ferroelectric materials are uniquely suited for this dual-mode operation, as their non-centrosymmetric crystal structures inherently exhibit coupled piezoelectric and pyroelectric responses. While traditional ceramic oxides and polymers have dominated this landscape, the push toward flexible, lead-free, and bio-compatible electronics demands an entirely new class of high-performance organic ferroelectric systems.[4]

In particular, conventional ferroelectric polymers such as poly(vinylidene fluoride) (PVDF), its copolymer poly(vinylidene fluoride-co-trifluoroethylene) [P(VDF-TrFE)], and odd-numbered nylons rely heavily on a long-range ordered crystalline structure to exhibit piezo-, pyro-, and ferroelectric properties.[5,6] Furthermore, the inherent structural polymorphism of these systems introduces significant processing bottlenecks; achieving the desired electroactive phases typically necessitates extreme conditions, such as mechanical pressures (~GPa) and intense electrical poling fields (~MV/m), which severely complicate device fabrication.[7,8]

In this regard, croconic acid ($C_5O_5H_2$) has attracted growing interest as a reference organic ferroelectric material, thanks to its unique combination of molecular structure, hydrogen-bonded crystal packing, and spontaneous polarization at ambient conditions. Croconic acid (CA) is a cyclic oxocarbon acid, belonging to the family of polyoxocarbon compounds, and is characterized by a conjugated system of oxygen-rich carbon rings. The molecular structure of croconic acid enables hydrogen bonding and proton transfer.[9,10] In contrast to traditional ferroelectric polymers or ceramic oxides, croconic acid forms a polar crystal phase stabilized by a dense network of hydrogen bonds, enabling long-range dipole ordering and reversible polarization switching.[11,12,10] Its spontaneous polarization, reaching ~20 μC/cm$^2$,[12] is comparable to or even surpasses that of conventional organic ferroelectrics, PVDF (~ 5-8 μC/cm$^2$)[13] and PVDF-TrFE (~ 9-12 μC/cm$^2$).[8] Moreover, its low coercive field (11-29 kV/cm) and molecular flexibility allow efficient domain switching under moderate electric fields.[14] The ferroelectric response of croconic acid is governed by a proton transfer mechanism along its hydrogen-bonding chains, which gives rise to a sharp phase transition and a strong dielectric signature.[12,15] These features make it suitable for a wide range of photonic and electronic applications, including capacitors, non-volatile memories, non-linear optics, low-power logic circuits, and sensors.[16–18] Progress in thin-film growth and crystallization control has also improved compatibility with nanoscale device structures requiring uniform polarization over large areas.[19] Beyond its well-studied ferroelectricity, CA also shows promise as a pyroelectric material due to the sensitivity of its polarization to temperature changes. Since the proton transfer mechanism is thermally activated, even moderate thermal variations can lead to measurable changes in polarization, making croconic acid a potential candidate for pyroelectric sensing and energy-harvesting applications, particularly on flexible, sustainable platforms. While quantitative data on its pyroelectric coefficients remain limited, several studies have reported temperature-dependent polarization behavior consistent with pyroelectric activity, particularly around the ferroelectric transition.[11,12] Croconic acid performance is closely linked to its degree of crystallinity. Highly ordered crystals support efficient dipole alignment and proton transfer, while structural disorder and defects tend to suppress ferroelectric functionality.[12,14] Consequently, the properties of polycrystalline croconic acid remain insufficiently reported, as most available data originate from single crystals with well-oriented hydrogen bonds. Therefore, the scalable fabrication of robust and large-area devices remains a critical bottleneck for practical energy harvesting applications with single crystals. Nevertheless, recent studies suggest that functional polarization can persist in polycrystalline films. For instance, croconic acid thin films, though polycrystalline, maintain room-temperature ferroelectricity, with polarization domains often matching grain morphology.[20] In addition, doping croconic acid into triglycine sulfate (TGS) has been shown to enhance pyroelectric response, an effect attributed to its large intrinsic dipole moment.[21]

In this work, we demonstrate the straightforward integration of vacuum-sublimated, room-temperature deposited continuous ferroelectric polycrystalline croconic acid films as active layers in pyroelectric and piezoelectric nanogenerators. The performance of the films is evaluated and benchmarked against state-of-the-art polymeric and hybrid nanomaterials through direct implementation in piezoelectric and pyroelectric nanogenerator layouts, achieving milestone piezoelectric and pyroelectric coefficient values for as-grown films without additional poling by tailoring CA thicknesses. We take advantage of post-encapsulation by a conformal adamantane dielectric polymer, deposited via remote plasma-assisted vacuum deposition at room temperature. The primary function of this conformal dielectric layer is to protect the polycrystalline films by suppressing surface-driven degradation processes, thereby ensuring long-term stability and enabling the fabrication of robust multilayered devices. However, its role becomes critical in optimizing device performance as a dielectric interface between the electrodes. We investigate the structural and functional properties of CA films, the electrical performance of the resulting devices, and the potential of these films as sustainable active materials for future piezoelectric, pyroelectric, and ferroelectric energy-harvesting applications.

## 2. Experimental Methods

### 2.1 Materials

Croconic acid powder (purchased from TCI, ≥ 98 % of purity, CAS: 488-86-8) and adamantane powder (purchased from Sigma Aldrich, ≥ 99 % of purity, CAS: 281-23-2) were used as received. Indium tin oxide/polyethylene terephthalate (ITO/PET) flexible substrates (obtained from Sigma-Aldrich, surface resistivity ~60 Ω/sq) were employed as electrodes. Adhesive copper foil (3M$^{TM}$ USA, nominal resistivity ~ 0.005 Ω/sq) was used to facilitate electrical connections. Silicon wafers (Topsil, <100> orientation, resistivity >10 kΩ·cm, thickness ~ 500 µm) served as the reference substrate for thin film deposition and were used for cross-sectional thickness characterization. The Alpha Wire 3049 series was used for the electrical connections in the devices.

### 2.2 Vacuum fabrication of croconic acid thin films and remote plasma encapsulation

Firstly, ITO/PET substrates underwent a standardized cleaning protocol comprising sequential immersion and ultrasonic agitation in anhydrous ethanol (99.9% purity) prior to the thin-film deposition. A site-selective chemical etching process was performed on the designated area to mitigate electrical short-circuiting in the fabricated device. A precisely dimensioned Kapton mask was used to define the deposition area, preventing polymer deposition beyond the intended boundaries and ensuring unobstructed electrical access to the ITO layer for contact formation.

Croconic acid films were obtained by thermal vacuum sublimation of CA powder at $10^{-6}$ mbar. The sample holder was kept at room temperature during the entire process. In situ thin film monitoring was performed using a quartz crystal microbalance (QCM) positioned adjacent to the sample holder. Subsequently, the films were encapsulated with a conformal adamantane plasma polymer film (hereafter ADA films), deposited by Remote Plasma-Assisted Vapor Deposition (RPAVD) in the same reactor. The RPAVD methodology is described in detail in previous references.[22–24] In summary, it involved the precise introduction of adamantane precursor vapor from a heatable dispenser into the downstream region of a 2.45 GHz, electron cyclotron resonance microwave (ECR-MW) Ar plasma, operated at 210 W at $2 \times 10^{-2}$ mbar. Argon gas was precisely regulated by a calibrated mass flow controller.

Prior to the CA thermal evaporation, the substrates were pretreated with a 150 W ECR-MW Ar plasma for 15 min at $5\times10^{-3}$ mbar in the deposition chamber. This plasma treatment was intended to clean and activate the surface to promote the formation of continuous CA films via vacuum sublimation.

### 2.3 Fabrication of croconic acid thin film based piezo and pyro-electric nanogenerator

The ADA-encapsulated CA thin films were deposited on ITO/PET pieces, which served as flexible, conductive substrates for fabricating piezo- and pyroelectric Croconic acid nanogenerators (hereafter, CA-NGs). The layout for the devices follows previous designs reported for thin film and low-dimension piezoelectric and pyroelectric nanogenerators.[25–27,9,8,13] The top electrode was formed by metal evaporation through a mask. For this, an aluminum foil mask, patterned with the electrode geometry, was fabricated using laser ablation (Explorer®One, 355 nm wavelength). Following precise alignment of the mask, gold/titanium (Au ~80 nm and Ti ~2-3 nm) electrodes were deposited by thermal evaporation under high-vacuum conditions (~$10^{-6}$ mbar). The substrates were maintained at 18 °C using a water-cooled sample holder during the evaporation process. The deposition rate and final thickness were monitored in situ using a QCM. Finally, for device assembly, electrical connections were established by attaching copper wires to the bottom (ITO/PET) and top (Au) electrodes using conductive silver epoxy and copper tape.

## 2.4 Characterization

The surface topography and piezoresponse force microscopic properties of the CA film were characterized using atomic force microscopy (AFM) with a Park NX-10 instrument. To ensure optimal sensitivity and conductivity, a platinum-iridium (Pt/Ir) coated conductive tip with a spring constant (k) of 3 N/m and a resonant frequency of 75 kHz was employed. Freshly deposited CA films on ITO/PET without protective layers were measured. The energy-harvesting performance of the CA-NGs, in terms of open-circuit voltage ($V_{OC}$) and short-circuit current ($I_{SC}$), was measured using a Keithley 2635A source meter unit.

For the piezoelectric characterization of the CA-NGs, a magnetic shaker (smart shaker K2007E01, Modal Shop) was used to impart precise kinetic activation, coupled with a function generator. The mechanical force imparted by the machine was calibrated using a force sensor (IEPE model 1053V2 from Dytran Instrument, Inc.) and recorded on a digital storage oscilloscope (Tektronix TDS1052B). Additionally, for the matching impedance figures of merit, the shaker was mechanically coupled to a cantilever beam on which the device was attached. The shaker applied a harmonic excitation to the cantilever, which underwent flexural deformation at its resonance frequencies. As the device followed the bending of the cantilever, it experienced cyclic mechanical strain, thereby generating a piezoelectric response. Further experimental details are provided in [28,29].

To evaluate the pyroelectric response of the CA-NGs, a heating gun, combined with an ice-cooled airflow, was used to impose controlled temperature oscillations via rapid, periodic switching between cooling and heating.[30] A thermocouple positioned near the device enabled real-time synchronization between the temperature stimulus and the electrical measurements, which were recorded using an oscilloscope. Data analysis and calculations were performed using the NanoDataLyzer software, available online.[31–33]

X-ray diffraction (XRD) patterns were recorded using a Panalytical X'PERT PRO diffractometer with Cu Kα radiation (λ = 1.5418 Å). Measurements were performed in grazing-incidence geometry with a fixed incidence angle of 0.25° to enhance surface sensitivity. The divergence slit was adjusted to match the film dimensions, ensuring that the X-ray beam remained within the sample area. Diffraction patterns were collected over a 2θ range of 10–90° at a scan rate of 0.3° $min^{-1}$. Baseline correction was applied to the diffractograms to compensate for background contributions from the substrate.

High-resolution field-emission scanning electron microscopy (FESEM) images of the samples deposited on silicon wafers were acquired with a Hitachi S-4800 microscope operated at 2 kV. Optical transmittance spectra of the samples deposited on fused silica substrates were recorded in the 200–2500 nm wavelength range using a PerkinElmer Lambda 750 S UV–vis–NIR spectrophotometer. Variable-angle spectroscopic ellipsometry (VASE) measurements were performed using a Woollam V-VASE ellipsometer. The optical constants were obtained by fitting the spectra to a Cauchy model.

## 3. Results and Discussion

### 3.1 Vacuum fabrication of croconic acid polycrystalline thin films

Croconic acid films were deposited by controlled thermal sublimation in vacuum. Before deposition, the substrates were treated in situ with Ar plasma at room temperature for ~10 min to improve wettability and avoid CA island formation observed in vacuum-sublimated films at room temperature.[20] Si(100), fused silica, and ITO/PET were used as substrates, yielding, in all cases, homogeneous distributions of CA crystalline aggregates coating the substrates (see **Figure 1a**). The thicknesses of the CA films are easily tunable by increasing deposition time, with thicker films developing longer crystalline aggregates. The films were not subjected to any post-deposition thermal treatments. As with other organic thin films, the as-grown layers evolve under ambient conditions.[34–36] The hygroscopic nature of the CA can also contribute to such degradation.[37–39] The limited stability of the as-deposited sublimated CA films prevents their direct use in the fabrication of operationally stable devices.

Thus, the initially uniform population of nanoscale crystallites progressively coarsens into micrometer-sized crystals, thereby exposing the substrate surface as demonstrated in the SEM images in **Figure S1**. To prevent such deleterious evolution, which usually involves amorphization or grain-boundary diffusion, we have taken advantage of encapsulation with adamantane plasma polymer formed by remote plasma-assisted deposition.[24,41,42] RPAVD adamantane films have previously been used as high-performance dielectric, transparent protective coatings for supported organic nanostructures,[24] as water and moisture protection for perovskite solar cells,[41] and as a flexible encapsulant for 2D materials for strain engineering.[42] A striking advantage of this plasma polymer is that it provides effective environmental and humidity protection at thicknesses below 100 nm, while maintaining high conformality. Moreover, the all-vacuum deposition of both the croconic acid and encapsulant layers allows for sequential formation in the same reactor, without exposing the interface to air or solvents. These favorable conditions yield highly stable configurations, showing neither morphological evolution nor crystalline coarsening even after ~18 months, as shown in **Figure S2** for adamantane thicknesses of ~500 nm and ~2μm.

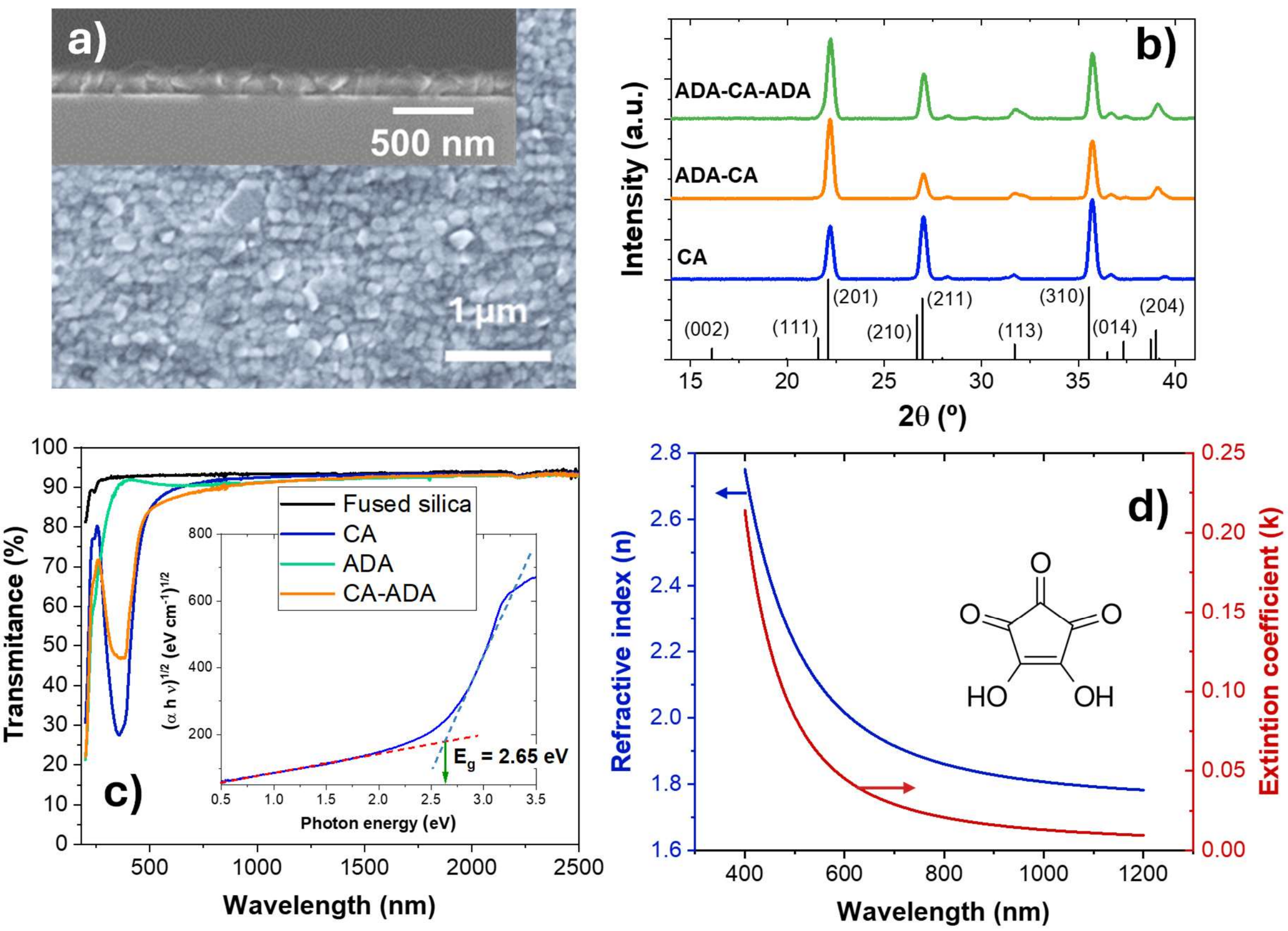

**Figure 1**. Characterization of the CA films. a) FE-SEM cross-sectional and planar micrographs of the surface of a freshly deposited CA film on Si(100). b) XRD pattern of encapsulated CA-ADA and as-grown CA films on Si(100), including the CA reference diffraction pattern.[40] c) UV-Vis-NIR transmission spectra of CA and ADA films deposited on fused silica. The inset shows the band-gap calculation of the CA film. d) Refractive index and extinction coefficient of a vacuum-sublimated CA film determined by VASE.

The croconic acid crystallizes in the orthorhombic polar space group $Pca2_1$ (no. 29), in which the ferroelectric polarization arises along the [001] direction due to cooperative proton transfer within the hydrogen-bond network.[10,12,40] **Figure 1b)** shows the grazing incidence XRD patterns of as-grown CA and CA-ADA films together with the reference diffraction pattern of crystalline CA. The spectra were acquired several hours after deposition. It is worth noting that adamantane encapsulation does not alter the crystallinity of the film. Indeed, XRD patterns for the encapsulated CA-ADA films exhibit high peak intensities and peak distributions that closely match those of the reference patterns. This result is in agreement with the time stability shown in **Figure S2** and serves as the basis for the nanogenerators investigated in this work. The (201) reflection dominates the diffraction pattern of the CA-ADA films, while other peaks, such as (211), (113), (310), and (204), also exhibit relatively high intensities, in some cases reaching about half of the (201) peak intensity. This distribution of reflections evidences the polycrystalline nature of the film, with several crystallographic orientations contributing to the diffraction pattern. Since the ferroelectric polarization lies along [001], the predominance of the (201) reflection indicates that the polar axis forms an oblique angle with respect to the film surface, implying that a component of the polarization lies within the plane of the film. The presence of other relatively intense reflections further confirms the coexistence of grains with different orientations. The XRD pattern of the as-grown CA films in **Figure 1b)** also reveals a polycrystalline structure, although with some differences in peak intensities. In this case, the most intense reflections correspond to the (211) and (310) planes, together with a significant contribution from the (201) peak, indicating a different distribution of grain orientations.

**Figure S3a)-b)** shows the time evolution of the grazing-incidence XRD patterns of the CA and ADA encapsulated CA films during the first 14 days after deposition. The diffraction patterns of the ADA-CA encapsulated films remain stable throughout this period, with no significant changes in peak positions, intensities, or shapes. In contrast, the XRD patterns of the CA film undergo a marked evolution, with new diffraction peaks appearing after six days together with additional low-intensity features. This behavior is likely associated with structural changes observed in unprotected CA films upon exposure to ambient conditions, leading to the formation of larger crystalline aggregates, as shown in **Figure S1**. By contrast, the ADA encapsulation preserves the initial distribution of the CA polycrystals formed at room temperature, thus preventing surface diffusion and interactions with atmospheric moisture and stabilizing the crystalline structure after deposition (**Figure S2**).

**Figure 1c)** shows that the vacuum-deposited films exhibit a strong absorption band centered at ~365 nm characteristic of the CA molecular absorption and a direct optical band gap of 2.65 eV (see the Tauc plot in the Fig.1c inset). The band gap is red-shifted relative to the reported calculated value of 3.2 eV for CA crystals.[18] This shift is very likely due to structural disorder and grain boundaries in the polycrystalline film deposited at room temperature. The films are non-dispersive and transparent at wavelengths $\lambda > 490$ nm, with slight absorption in the 400-490 nm range responsible for the faint yellowish coloration. The refractive index of the CA films reaches relatively high values in the visible region due to absorption in the initial part of the visible, yielding a value of n ~1.8 in the NIR region (**Figure 1d**) for as-grown samples.

Panel c) also compares with the transmittance for ADA and CA-ADA deposited on fused silica. The transmission spectra of the CA-ADA film match those of the non-encapsulated films because the adamantane plasma polymer is fully transparent in the visible range, enabling the production of transparent encapsulated devices.[24]

### 3.2 Nanoscale electromechanical properties of CA thin film

Atomic force microscopy (AFM) was employed to investigate the surface topography of freshly deposited organic polycrystalline CA thin film (thickness, t ~150 nm). **Figure 2**a) presents the AFM surface topography of the CA thin film, with a typical average surface roughness of $S_{RMS}$ ~13 ± 2 nm. A closer view of the topography **(Figure 2b)** of the marked region A unveils the preferential growth of partly elongated needle-like structures. These elongated structures exhibit a typical length of ~100 ±25 nm and a width of ~20 ± 8 nm. The preferential growth suggests an anisotropic crystallization process, potentially influenced by the directional molecular assembly of CA.[43] Notably, some plateau-like features with slightly elevated heights are also present in the film (marked in **Figure 2a**) that can be attributed to substantial diffusion of croconic acid molecules on the surface.

Furthermore, the localized switching behavior of piezoelectric and ferroelectric dipole orientations within the CA thin film was investigated by piezoresponse force microscopy (PFM). This technique enables mapping piezoelectric and ferroelectric activity at the nanoscale. A schematic representation of the PFM measurement is illustrated as an inset in **Figure 2f)**. To obtain the PFM signal, a sharp conductive tip (Pt/Ir) was brought into contact with the sample surface, and then an AC voltage of amplitude $V_0$ ~5 V was applied between the conductive tip and the conductive substrate (ITO). By scanning the tip across the surface, a map of the piezo (amplitude) and ferro (phase) responses can be generated by applying an AC signal through the tip to the samples during the scan in contact-mode PFM.[5] We employed both vertical (VPFM) and lateral (LPFM) PFM configurations to observe the induced amplitude and phase deflections in response to vertical and shear stimuli, respectively.

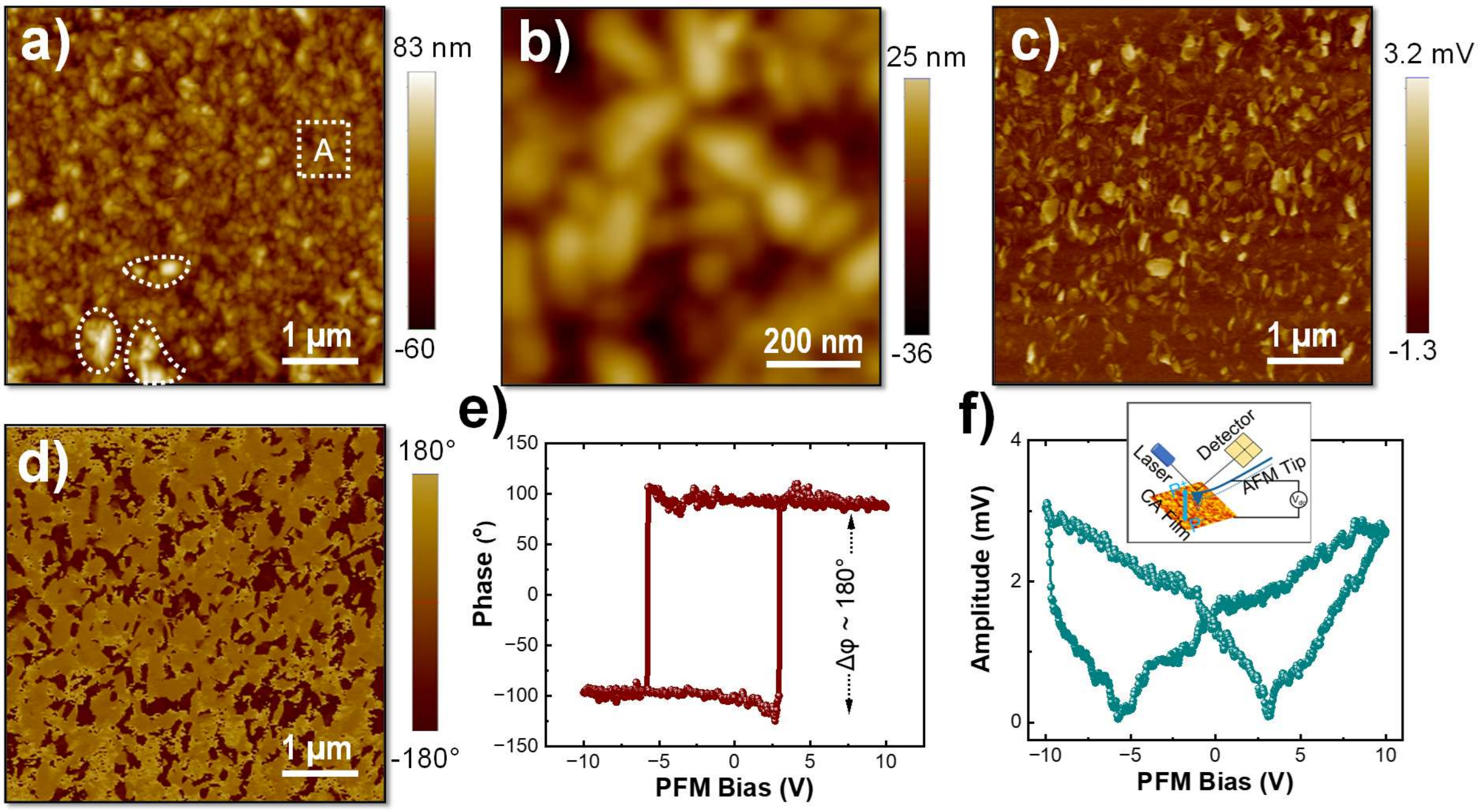

**Figure 2.** Piezoresponse force microscopy characterization and local ferroelectric switching of CA thin films. a) Surface topography of CA thin film. b) High-resolution topography of the marked region "A" revealing distinct needle-like crystalline structures. Corresponding PFM map showing the local c) amplitude and d) phase contrast variations across the surface. PFM switching spectroscopy loops acquired on the film: e) a sharp, rectangular phase hysteresis loop confirming 180º ferroelectric polarization switching, and f) a characteristic butterfly-shaped amplitude loop demonstrating piezoelectric nature and schematic illustration of the AFM/PFM experimental measurement configuration in the inset.

The VPFM amplitude and phase images (**Figure 2c-d**) reveal regions of distinct piezoelectric activity, indicating the presence of polar domains within the CA thin film. The observed phase contrast suggests that these domains exhibit both upward ($P^+$) and downward ($P^-$) polarization switching. Localized VPFM hysteresis loops (**Figure 2e-f**) exhibit complete ferroelectric-like behavior and $\Delta\varphi \approx 180°$ dipole orientation reversal and a clear butterfly-shaped loop under the application of an external DC bias voltage. Notably, a very low coercive field ($E_C$ ~ 194 kV/cm) is observed in our system, compared to the previously reported $E_C$ value for quasi-2D polycrystalline croconic acid films deposited at low temperature on $Al_2O_3$ by high-vacuum PVD.[20] This remarkably low $E_C$ value is strong evidence of the superior electroactive nature of the CA films, suggesting a substantial piezoelectric and ferroelectric response. In addition, we have performed PFM reading and writing for localized domain switching in an uncapsulated croconic acid thin film (~100 nm) using the box-in-box method. Initially, a downward polarization state ($P_{down}$) was written over a 5×5 μm$^2$ region by applying a +10 V DC bias to the conductive AFM tip. Subsequently, a 3×3 $\mu m^2$ area was scanned by applying a -10 V DC bias to switch the polarization ($P_{up}$), thus effectively erasing the pre-written domain. Following this, a large area (12×12 $\mu m^2$) was scanned without any DC bias to read out domain switching. The out-of-plane PFM phase image (**Figure 3a**) and its corresponding line profile (**Figure 3c**) reveal sharp, well-defined 180° contrast boundaries between the unwritten region, written ($P_{down}$), and erased ($P_{up}$) domain states, confirming typical robust domain confinement and non-volatile charge retention characteristics. Simultaneously, the PFM amplitude image (**Figure 3b**) and its drawn line profile (**Figure 3d**) display uniform electromechanical activity within the poled region. This characteristic offers a promising pathway for miniaturizing ferroelectric non-volatile memory devices and energy harvesting applications toward flexible and sustainable organic electronics.

In contrast, well-known ferroelectric polymers such as PVDF and PVDF-TrFE exhibit significantly high $E_C$ values, typically ranging from 400 to 1000 kV/cm, depending on processing conditions.[44,45] However, these polymers also possess inherent processing and structural limitations and require complex optimization for stabilizing the highly desired electroactive phase in β-PVDF. However, optimizing the preferred electroactive β-phase in PVDF-TrFE is further constrained by its low ferroelectric-to-paraelectric transition temperature ($T_c$), which exhibits a strong compositional dependence on the molar fraction of TrFE units in the PVDF matrix. Additionally, both polymers typically rely on solution-based processing for thin-film fabrication. In contrast, croconic acid thin films offer a compelling alternative with a robust coercive field compared to conventional PVDF-based polymers, and benefit significantly from solvent-based deposition and from bypassing phase-stabilization constraints.

Furthermore, amplitude hysteresis (butterfly loop) demonstrates the lattice expansion and contraction within the crystal domain under ±10 V bias. The effective piezoelectric coefficient ($d_{33}$) in VPFM mode, calculated from the ratio of the change in amplitude (ΔZ) to the applied bias (ΔV), yields a value of ~15 pm/V.[5] Remarkably, the obtained $d_{33}$ value is very close to those of well-established organic polymers, such as PVDF and PVDF-TrFE.[46–48] We employed LPFM to probe the in-plane piezoelectric and ferroelectric properties of the CA thin film, in conjunction with VPFM data. LPFM is particularly sensitive to shear deformations and provides insights into the polarization components parallel to the film surface, as schematically demonstrated in **Figure S4a)**. The amplitude image (**Figure S4b**) reveals regions of varying piezoresponse, suggesting a non-uniform distribution of in-plane piezoelectric activity. Some elongated features of large amplitude are observed, with a characteristic lateral length $l$ ~100 ± 25 nm, which appear to align with the elongated structures observed in the AFM surface topography (**Figure 2a**). The phase image (**Figure S4c**) exhibits distinct contrast, with regions of 180° phase shifts, indicating the presence of antiparallel in-plane polarised domains. In this connection, the phase change and butterfly spectroscopic response demonstrate the typical ferroelectric and piezoelectric characteristics of the in-plane geometry (**Figure S5**). This suggests that the CA thin film possesses a well-defined in-plane ferroelectric polarisation, which is essential for engineering materials with a large pyroelectric response,[49] since the secondary part of pyroelectricity depends on the in-plane piezoelectricity. Thus, vacuum-deposited CA thin films are promising candidates for large-scale pyroelectric energy-harvesting systems.

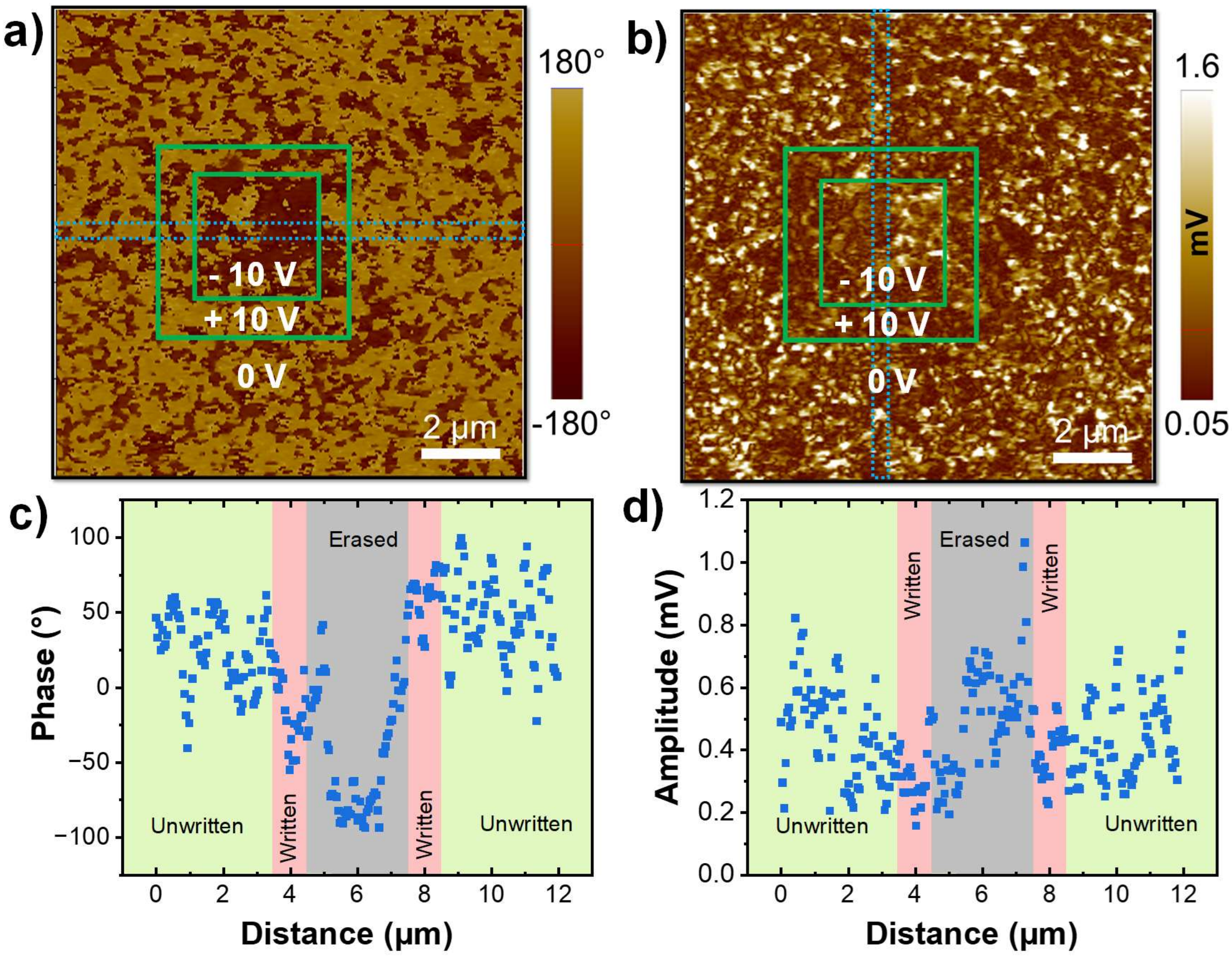

**Figure 3.** Nanoscale ferroelectric domain switching characteristics of CA thin film. a) PFM phase and b) amplitude maps demonstrating domain writing and erasing response in out-of-plane configuration. Box-in-box pattern highlights regions under +10 V (written), -10 V (erased), and 0 V (unwritten) electrical poling. C, d) Corresponding line profiles extracted across the dashed blue lines in (a) and (b), respectively. The profile in c) shows distinct phase-contrast shifts mapping across the unwritten, written, and erased regions, confirming successful

180º polarization switching. The profile in d) tracks variations in the local piezoresponse amplitude across domain walls and written/erased boundaries.

### 3.3 Piezoelectric energy harvesting performance

The previous section established the ferroelectric and piezoelectric responses of as-grown CA thin films at the nanoscale. However, their polycrystalline nature and limited environmental stability may hinder direct implementation in energy-harvesting devices. The first concern is mitigated by the results in previous reports showing that polycrystalline and texturized films, such as ZnO thin and thick films, can operate reliably in piezoelectric nanogenerators.[25–27] To address the second limitation, we propose using adamantane-encapsulated croconic acid films to fabricate piezoelectric nanogenerators. The first device configuration follows the designs in previous articles, using a layer-by-layer approach with a smaller top electrode and shadow-mask deposition to avoid short-circuiting at the boundary with the ITO bottom layer and to precisely define the active area of the nanogenerator.[50,25,27,29,30] The schematic and photograph of the devices with the wired terminals are presented in **Figure 4a)**. We label this device as CA(500 nm)-ADA(100 nm) to reflect the selected layer thicknesses. The nanogenerator multilayers were simultaneously deposited on ITO-coated PET substrates for device fabrication and on Si(100) substrates for cross-sectional SEM characterization. The cross-section SEM micrograph of **Figure 4a)** displays the stacked heterostructure comprising a granular croconic acid film covered by the adamantane plasma polymer encapsulation layer and an evaporated top electrode (Au/Ti), showing a well-defined layer-by-layer device in which the adamantane and top electrode films grow conformally over the CA nanocrystals. The nanogenerator was tested using an electromagnetic shaker applying a vertical excitation in a cantilever configuration. Upon this kinetic excitation source, the entire device is bent, imposing a periodic deformation on the crystalline macromolecular structure of the CA film. This deformation modulates the intrinsic piezoelectric polarization, generating an output voltage and driving a compensating piezocurrent through the external load [25,29] (see additional details in the Experimental section).

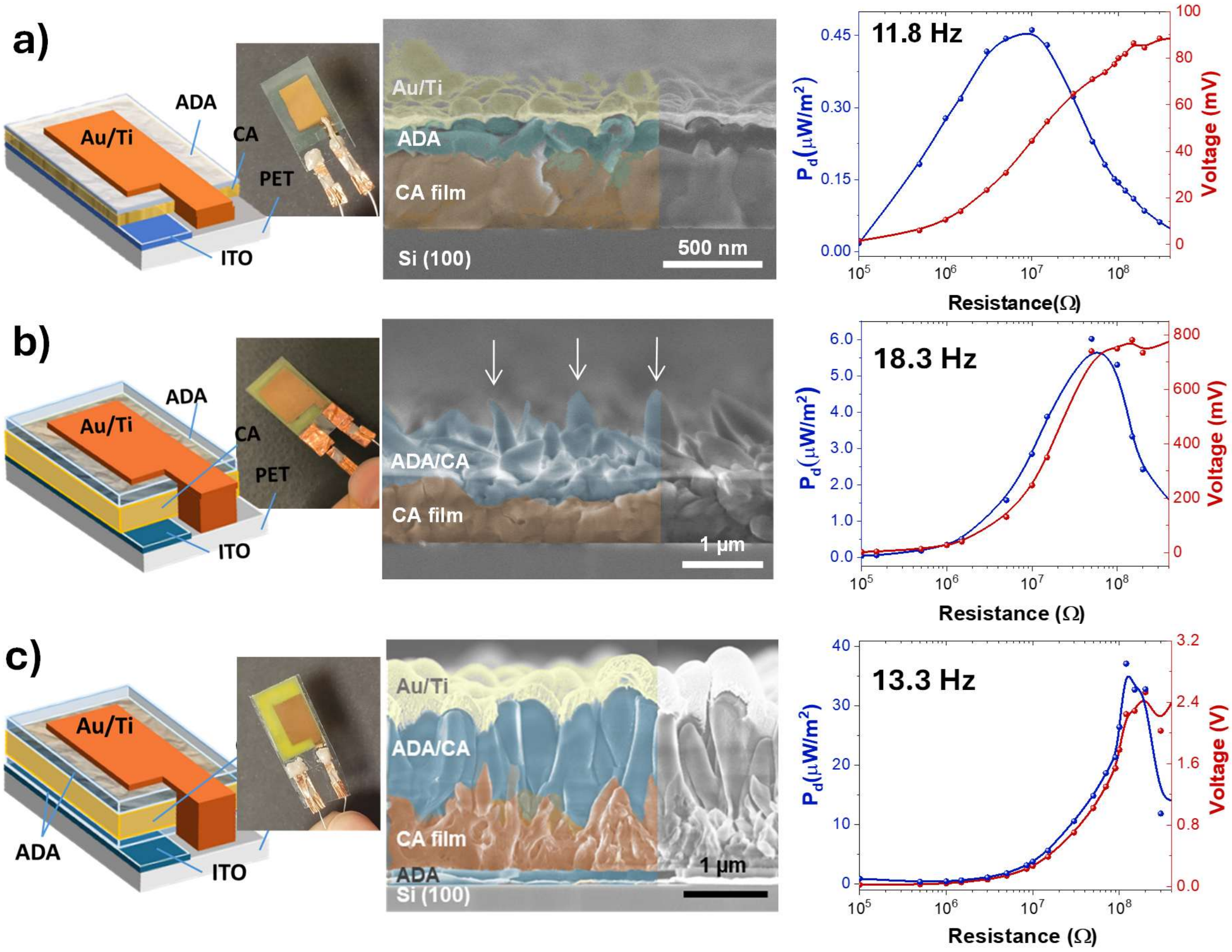

**Figure 4. Piezoelectric energy harvesting performance of CA-ADA-NGs.** Schematic of the device layout and photograph of the device (left), cross-sectional SEM micrograph (middle) and impedance matching characteristics of the output voltage with varying load resistances and corresponding output power density for optimal frequencies (right) for (a) CA(500 nm)-ADA(100 nm), (b) CA(1.6 μm)-ADA(200 nm), and (c) ADA(50 nm)-CA(2 μm)-ADA(1 μm). The reported values for output voltage and power density correspond to the optimal cantilever frequencies.

**Figure S6** shows the open-circuit ($V_{OC}$) and short-circuit current ($I_{SC}$) responses as a function of vibration frequency (see the characteristic open-circuit voltage ($V_{OC}$) and short-circuit current ($I_{SC}$) output vs. time curves presented in **Figure S7**). Meanwhile, the $V_{OC}$ shows a linear increase up to ~55 mV (at 20 Hz), the $I_{SC}$ values increase with frequency, reaching 174 nA/cm$^2$ at 15 Hz, and then remaining stable at 177 nA/cm$^2$ at 20 Hz for the CA(500 nm)-ADA(100 nm) device. The $I_{SC}$ plateau indicates a limitation in the linear response to rapid mechanical stimuli and suggests that charge carrier dynamics within the CA active film may be governed by intrinsic relaxation times or interfacial resistances.[51] Nevertheless, the measured $I_{SC}$ values remain notably high, highlighting effective conversion of mechanical deformation into electrical current. This observation features $I_{SC}$ as a key parameter for device performance, since higher $I_{SC}$ values lead to increased power output and improved functionality under practical conditions.[52,53] To evaluate the electrical output performance of the CA(500 nm)-ADA(100 nm) device, a systematic investigation of the voltage output ($V_L$) as a function of external load resistance ($R_L$) was conducted (**Figure 4a**) for a vibration frequency set at 11.8 Hz. The output voltage increases monotonically with increasing load resistance, reaching a saturation plateau at $R_L \sim 10^8$ Ω. The observed saturation can be attributed to impedance matching, which occurs when the external load resistance approaches the CA-ADA internal resistance, thereby maximizing voltage transfer and minimizing internal power dissipation.[54,55] Similarly, the mean power density ($P_d$) of the device was also determined as a function of the external load resistance and was calculated using the established formula: $P_L(t) = V_L^2(t)/(R_L \times A)$ where A represents the active area of the device. The analysis revealed a peak power density of 0.45 μW/m$^2$, which corresponds to the maximum electrical power the device can deliver to an external load under the specified operating conditions. This maximum power delivery is obtained at a matching impedance of 10MΩ. Note that this impedance-matching value is nearly two orders of magnitude below the saturation output-voltage value. This apparent discrepancy is a direct consequence of the capacitive nature and charge-transfer rate of the CA-NG device.[54,56]

The first results shown in **Figure 4a)** are encouraging for the implementation of CA-ADA bilayers in PENGs, however, the overall performance is still not competitive with other organic and polymeric piezoelectric systems. It is important to note that benchmarking across these systems remains challenging due to the lack of consensus on the excitation conditions or boundary constraints used to evaluate their figures of merit. Consequently, reported values often reflect different testing modes, device geometries, and mechanical loading conditions, which complicates direct comparison. It is worth noting that the performance of conventional piezoelectric nanogenerators is typically evaluated using vertical compression (direct pressing) in a top-bottom geometry. This approach often introduces artifacts and yields mixed responses due to triboelectric effects. To eliminate these ambiguities and isolate the true piezoelectric performance of the device, we used a cantilever-based mechanical excitation. Consequently, a direct comparison between our PENG device performance and values reported in the literature is not straightforward due to the varying testing methodologies. Despite this divergence, the summary in **Table S1** shows that the maximum peak power values for PVDF, PVDF-TrFE, and azobenzene derivatives already reach the μW/cm$^2$ range. Although the primary aim of this work is not device optimization but rather to demonstrate the proof-of-principle of vacuum-processed croconic acid layers, this comparison motivated us to undertake an initial attempt to improve the piezoelectric performance.

The first step to improve the device performance was to increase the CA thickness to ~1.6 µm. **Figure 4b)** shows the device layout and photograph, and a cross-sectional SEM micrograph. The increase in thickness also increased the surface roughness of the CA polycrystalline layer due to the formation of elongated grains (see arrows in the micrograph). Note that, as a result, the thicknesses of both the CA film (~1.6 µm) and the ADA, as determined from the SEM analysis, are only approximate. To ensure coverage and maintain continuity of the Au/Ti top electrode, the thickness of the adamantane layer was also increased relative to the first device up to ~200 nm, resulting in a CA(1.6 µm)-ADA(200 nm) nanogenerator. It should be noted that the boundary with the conformal ADA layer becomes less distinct in this configuration (**Figure 4b**). Thus, the reported ADA thickness corresponds to that of an ADA film deposited simultaneously on a reference Si (100) substrate. The corresponding piezoelectric output characterization reveals a significant enhancement in comparison with the thinner CA film counterpart. **Figure S8** shows the characteristic $V_{OC}$ and $I_{SC}$ vs. time curves, yielding values as high as 0.8 V peak-to-peak and 3.8 nA/cm$^2$ at an optimal cantilever frequency of 18.3 Hz. The maximum $V_L$ and $P_d$ values (**Figure 4b**) increase by more than an order of magnitude in the first case and by 15 times in the second.

On the one hand, that performance improvement can be attributed to fundamental differences in the dipole density and depolarization-field characteristics between the thin (~500 nm) and thick (~1.6 µm) CA films. The thicker CA film likely has a significantly higher concentration of permanent dipoles than the thin film, which directly translates into a greater capacity for charge separation and subsequent electrical output upon mechanical deformation. In addition, in thin CA films, the depolarization field, which opposes the polarization induced by mechanical stress, is substantial and scales inversely with thickness, thus significantly reducing the effective polarization and extractable power output.[57,58] Conversely, the thicker film contains sufficiently developed surface and polarization domains that mitigate this effect. This robust domain network effectively shields the bulk material from the depolarization field, thereby preserving a higher degree of polarization and maximizing electrical output. On the other hand, the introduction of dielectric or semiconducting interlayers between the active piezoelectric film and the electrode has been shown to reduce interfacial charge recombination in piezoelectric nanogenerators.[27,29] In our case, the estimated dielectric constant of the adamantane is $\varepsilon_r' \sim 2.67$.[24,59] In the comparison between CA(500 nm)-ADA (100 nm) and CA(1.6 μm)-ADA(200 nm), the higher adamantane thickness can also increase the open-circuit voltage through in-series capacitance reduction. Indeed, the implementation of an interlayer has also been reported to stabilize the polarization-induced piezo voltage by preventing charge leakage and screening effects at the electrode interfaces.[29,50,60]

Considering the previous factors, we fabricated a third piezoelectric nanogenerator that implements a double adamantane layer to interface the top and bottom electrodes. The layouts, photograph, and cross-sectional micrograph of the ADA(50 nm)-CA(2 μm)-ADA(1 μm) are presented in **Figure 4c)**. The croconic acid grows similarly on the adamantane layer as on the reference Si substrate, as indicated by the comparison of the cross-sectional micrographs in the figure. Moreover, the ADA-CA-ADA grazing incidence XRD patterns in **Figure 1b)** exhibit a polycrystalline signature similar to that of the CA-ADA films on Si(100), together with the same structural stability under ambient conditions (see **Figures S2a)-b)** and **S3b)-c**).

The output voltage of that device outperforms the previous one by almost one order of magnitude, reaching a maximum voltage of ~3 V and yielding $P_d$ as high as ~37 μW/m$^2$ (see **Figure 3c**), and corresponding $V_{OC}$ and $I_{SC}$ vs time curves in **Figure S9**. This optimization pathway is straightforward and works with vacuum-processable piezoelectric, ferroelectric, and dielectric counterparts. It is worth stressing at this point that the entire device is produced in compliance with the vacuum and plasma one-reactor premises, previously exploited for the development of nanoelectrodes, photoanodes, and ZnO-based piezoelectric nanogenerators.[25,26,61–63]

### 3.4 Pyroelectric energy harvesting performance

Inspired by the exceptional intrinsic piezoelectric and ferroelectric properties of CA, we sought to evaluate its response to temperature oscillations, i.e., its pyroelectric behavior. Because pyroelectricity is an inherent hallmark of true ferroelectricity, demonstrating a robust pyroelectric response is essential to validating the fundamental polar nature of the CA thin films. To date, no proof-of-concept pyroelectric device comprising CA thin films has been reported in the literature. Although, the piezoelectric output performance of our PENG devices is modest compared to benchmark ferroelectric materials, CA offers a profound processing advantage. Crucially, it does not require any post-deposition treatments, such as the complex poling or stretching protocols as required for benchmark PVDF and its copolymers to attain their electroactive phase and crystallinity. Consequently, its dual piezo/pyroelectric capabilities, combined with processing simplicity, position CA as an exceptionally efficient material system for future functional nanoelectronics.

Therefore, the phenomenon of pyroelectricity enables thermal energy harvesting by converting otherwise untapped waste-heat oscillations into usable electrical energy.[64,65] In this regard, a multilayered ADA (50 nm)-CA (2 µm)–ADA (1 µm) device (schematically illustrated in **Figure 5a)** was employed for pyroelectric thermal energy harvesting under temperature oscillations between ~300 K and 340 K, to demonstrate its potential as a robust platform for thermal energy harvesting and sensing. As an initial observation, the changes in spontaneous polarization ($P_S$) can be attributed to randomly oriented dipoles responding to temperature oscillations, thus confirming the pyroelectric behavior of the CA film. The applied thermal stimulus input consisted of a temperature fluctuation of $\Delta T$ ~40 K and a heat rate of dT/dt ~10 K/s (**Figure 5 b-c**) (see additional measurement details in the Experimental Methods Section and elsewhere.[30] This thermal input enabled the measurement of an $I_{SC}$ of ~10 nA (**Figure 5d**). This experimentally examined $I_{SC}$, analyzed further using the Lang and Stackle method,[66] yielded a pyroelectric coefficient (*p*) of 10 µC/m$^2$K, which substantiates the effectiveness of CA thin films in generating electrical signals in response to thermal variations. The pyroelectric coefficient was evaluated by using the formula $I_{SC} = pA dT/dt$, where A is the effective area of heat exposed for ADA-CA-ADA nanogenerator.[64] The experimentally observed values of $I_{SC}$ and *p* demonstrate that the ADA-CA-ADA multilayer structure exhibits stable and reproducible pyroelectric behavior under controlled thermal heating and cooling cycles. This stability is critical for the potential integration of such materials into functional devices. Notably, the estimated pyroelectric coefficient of the CA pyroelectric harvester lies within the expected range for efficient thermal-to-electrical energy conversion, positioning CA thin films as viable candidates for thermal sensing, infrared detection, and thermal energy harvesting. The comparative analysis shown in **Figure 5e)** benchmarks the performance of CA-NG devices against established pyroelectric materials, highlighting their competitive output and potential for advanced pyroelectric applications. [30,67–69,69–76]

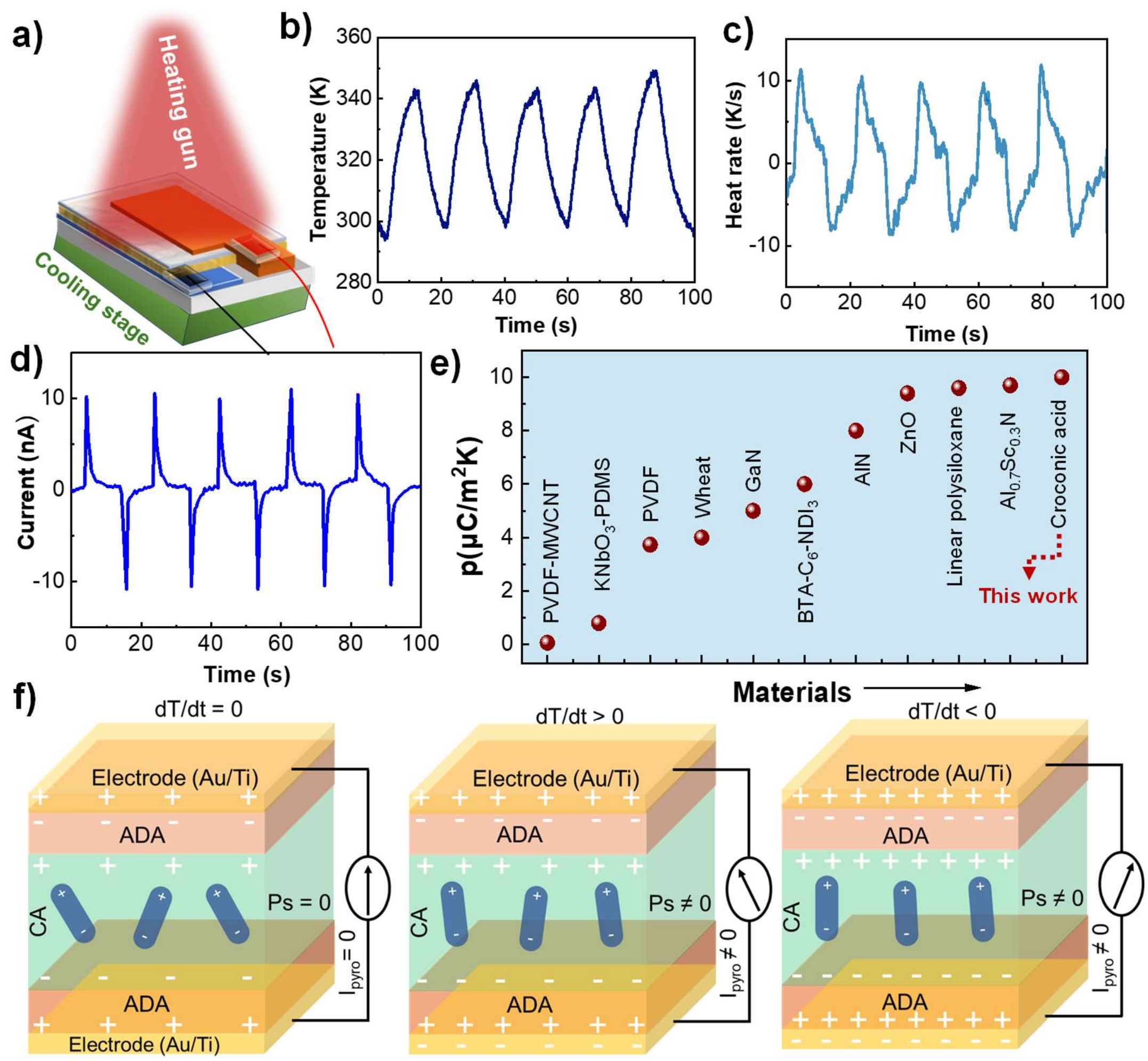

**Figure 5**. **Pyroelectric energy harvesting with ADA-CA-ADA nanogenerators.** a) Schematic of the experimental procedure to expose the ADA (50 nm)-CA (2 µm)-ADA (1 µm) NG to thermal fluctuation. b) Experimental periodic input on/off temperature profile (ΔT→ 40 K). c) Corresponding heat rate (dT/dt), d) measured short-circuit current ($I_{SC}$) under temperature oscillations, e) comparative analysis of the ADA-CA-ADA PyNG against state-of-the-art pyroelectric materials.[30,67–69,69–76] f) Proposed mechanism for pyroelectric signal generation in the CA-NG under thermal fluctuations.

The mechanism behind the pyroelectric signal generation in ADA-CA-ADA NGs can be summarized as follows. Pyroelectric materials exhibit spontaneous polarization $P_S$ due to the asymmetric distribution of dipoles within the crystal lattice. When subjected to a thermal gradient ($dT/dt \neq 0$), the polarization changes as ($p = dP_S/dt$).[64] At $dT/dt = 0$, the CA-NG reaches a steady state, and there is no net change in polarization. However, intrinsic domain relaxation and residual lattice stress may still induce minor fluctuations in charge distribution.

This behavior is attributed to thermally induced random oscillations of electric dipoles around their dipole alignment axis. As a result, the total average power of the electric dipole's spontaneous polarization will be constant, and there will be no electricity output (schematically demonstrated in **Figure 5f)**. Subsequently, when the temperature rises ($dT/dt > 0$), the increased thermal energy disrupts hydrogen bonds in the croconic acid molecules. Consequently, the dipole oscillations become more dynamic, reducing the net polarization. The decrease in polarization leads to a change in the surface charge, causing electrons to flow as a pyroelectric signal (**Figure 5f**). When $dT/dt < 0$, the decrease in thermal energy leads to a corresponding reduction in dipole oscillations. This allows the dipoles to align more effectively, increasing the net polarization and leading to a rise in the charges induced on the electrodes (**Figure 5f**). Consequently, induced electrons flow in the opposite direction compared to the heating cycle. This process effectively optimizes the pyroelectric response of the CA-NG device. As with piezoelectric characterization, interface engineering and layer-thickness optimization can be further pursued to improve the device performance under realistic thermal stimuli.

## Conclusions

Croconic acid films are deposited on Ar plasma-treated surfaces via vacuum sublimation, yielding continuous polycrystalline layers. To prevent degradation upon exposure to air, the films were encapsulated in situ with an adamantane remote plasma polymer. This encapsulation ensures long-term stability under ambient conditions for more than one year and effectively suppresses the surface crystallization processes observed in uncoated polycrystalline CA films. The resulting polycrystalline CA films exhibit ferroelectric behavior, with the polarization axis oriented at an oblique angle relative to the film surface. PFM analysis confirms the presence of well-defined polar domains that exhibit complete ferroelectric switching, with a relatively low coercive field ($E_c \sim 194$ kV/cm).

Leveraging their ferroelectric and piezoelectric properties, the encapsulated CA films were implemented as active layers in multilayered piezoelectric and pyroelectric CA-NGs. Piezoelectric characterization reveals that the output mean power density increases with film thickness and with the presence of a double adamantane interface, reaching ~ 37 $\mu W/m^2$ for ~1 μm-thick films. Regarding pyroelectric energy harvesting, the ADA-CA-ADA NGs demonstrate state-of-the-art performance with pyroelectric coefficients of up to 10 $\mu C/m^2K$. These values surpass those of widely used pyroelectric materials such as ZnO, AlN, GaN, and polymer-based systems, including PVDF, highlighting the strong potential of this platform for advanced thermal energy harvesting and sensing applications. The results indicate that encapsulated, vacuum-sublimated polycrystalline CA films have strong potential for multi-energy scavenging. Beyond the achieved performance, the proposed fabrication route offers significant advantages over conventional methods, including solvent-free processing, low-temperature compatibility, and fully in situ integration of film deposition and encapsulation. This approach ensures clean interfaces, reduces process complexity, and improves reproducibility, while remaining compatible with flexible substrates and scalable vacuum-based technologies. The simplicity of the fabrication procedure is particularly attractive for developing practical devices, offering ample room to optimize deposition parameters and paving the way for lead-free, low-cost, fully organic, high-performance energy-harvesting systems.

*Acknowledgements*

We thank the projects PID2022-143120OB-I00, PCI2024-153451 funded by MCIN/AEI/10.13039/501100011033 and by "ERDF (FEDER) A way of making Europe". Project ANGSTROM was selected in the Joint Transnational Call 2023 of M-ERA.NET 3, which is an EU-funded network of about 49 funding organizations (Horizon 2020 grant agreement No 958174). The project leading to this article has received funding from the EU H2020 program under grant agreement 851929 (ERC Starting Grant 3DScavengers). The authors also want to thank the CSIC Interdisciplinary Thematic Platform (PTI) Susplast. GPM acknowledges a FPI Grant from the MINECO/AEI Program PRE2020-093949.

## Supporting Information

## Polycrystalline ferroelectric croconic acid for multisource environmental energy harvesting

*Gloria P. Moreno-Martínez, Hari K. Mishra,* X. García-Casas, María Alcaire, Vanda Godinho, Juan R. Sánchez-Valencia, Francisco J. Aparicio, Ana Borrás, Angel Barranco**

Nanotechnology on Surfaces and Plasma Group, Materials Science Institute of Seville (CSIC-US), c/Américo Vespucio 49, 41092 Seville, Spain

***e-mail:** hari.krishna@icmse.csic.es; angel.barranco@csic.es

**Evolution of unprotected CA and ADA encapsulated films stored in air.**

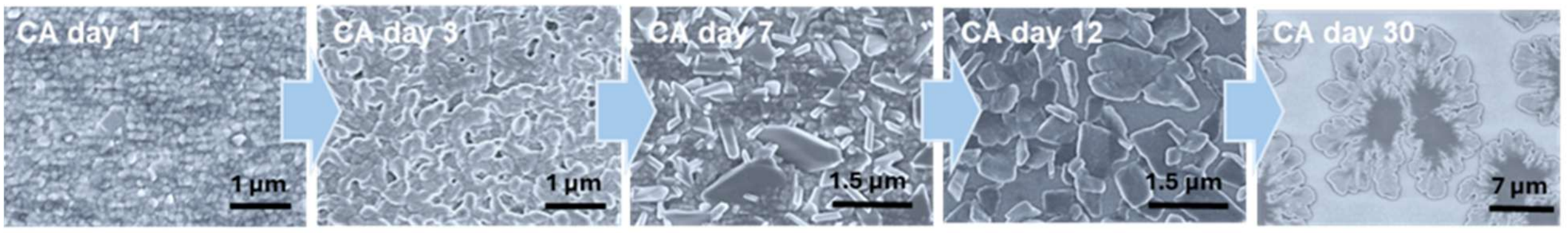

**Figure S1**. Morphological evolution of the polycrystalline structure of unencapsulated, vacuum-sublimated CA films during a period of 30 days of storage under ambient conditions (~50-55% relative humidity and ~22-27 ºC).

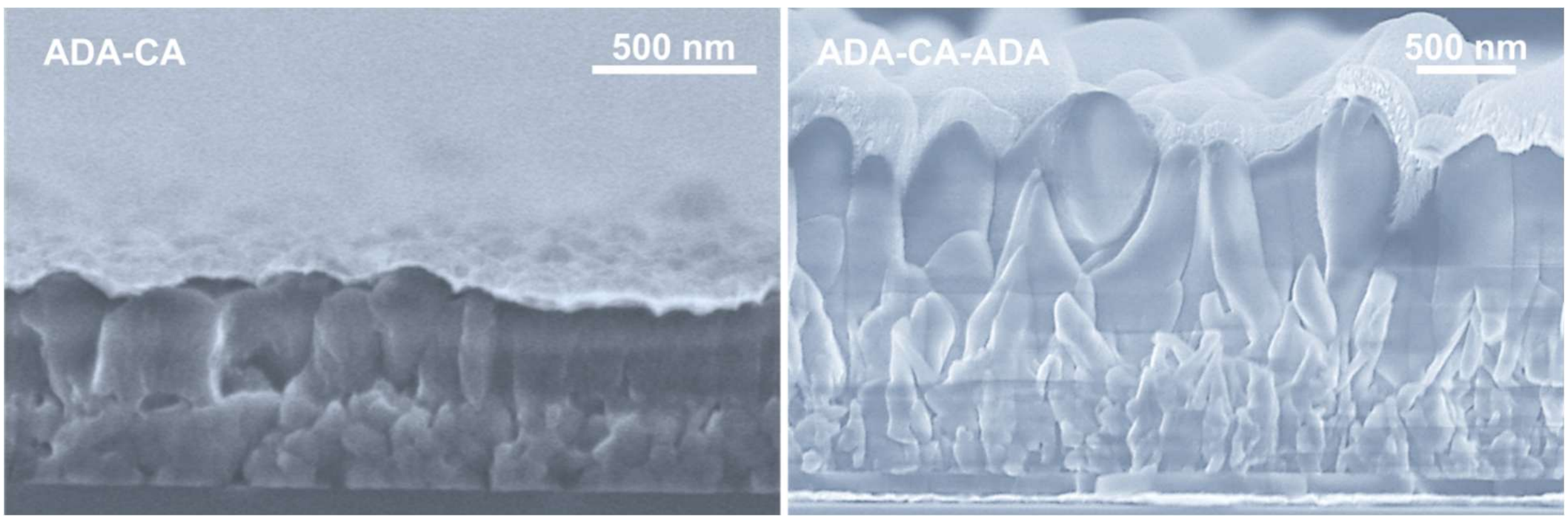

**Figure S2**. Multilayered encapsulated CA-ADA films of different thicknesses after ~18 months of storage under ambient conditions.

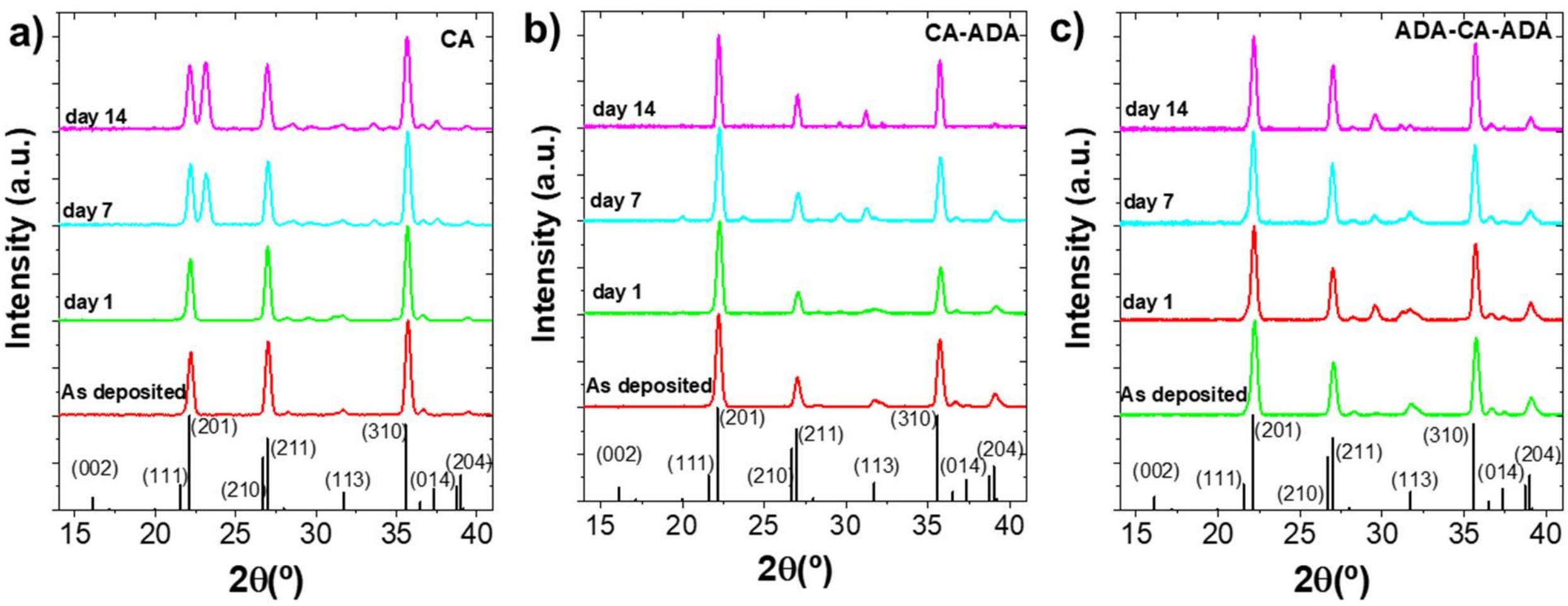

**Figure S3**. Evolution of grazing incidence XRD diffraction vs. time for a) CA, b) CA-ADA, and c) ADA-CA-ADA systems deposited on Si(100) and stored under ambient conditions.

**Lateral PFM characterization**

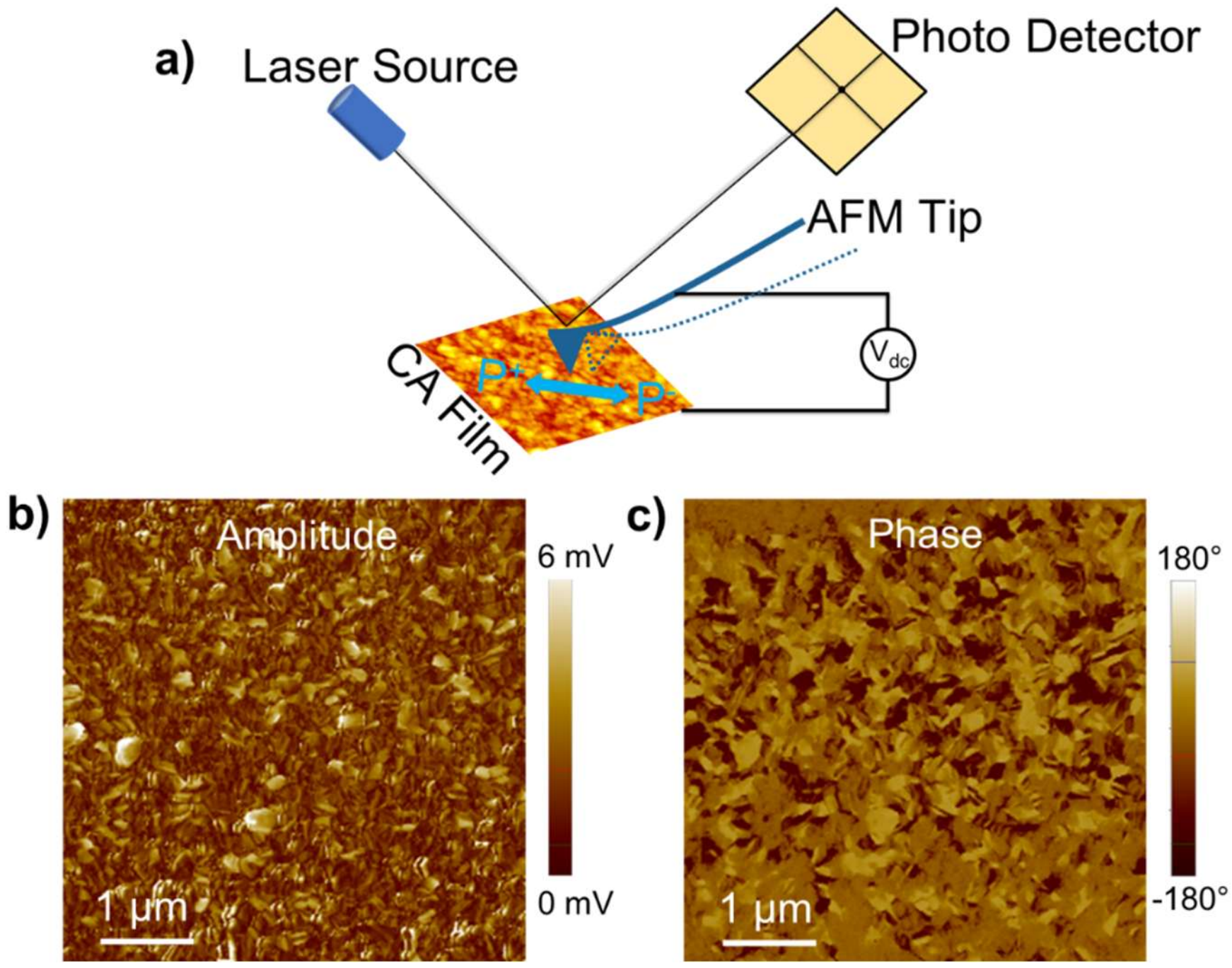

**Figure S4**. a) Schematic demonstration of LPFM measurement and corresponding induced polarization in the in-plane configuration. b) Change in amplitude and c) phase maps, under LPFM scan.

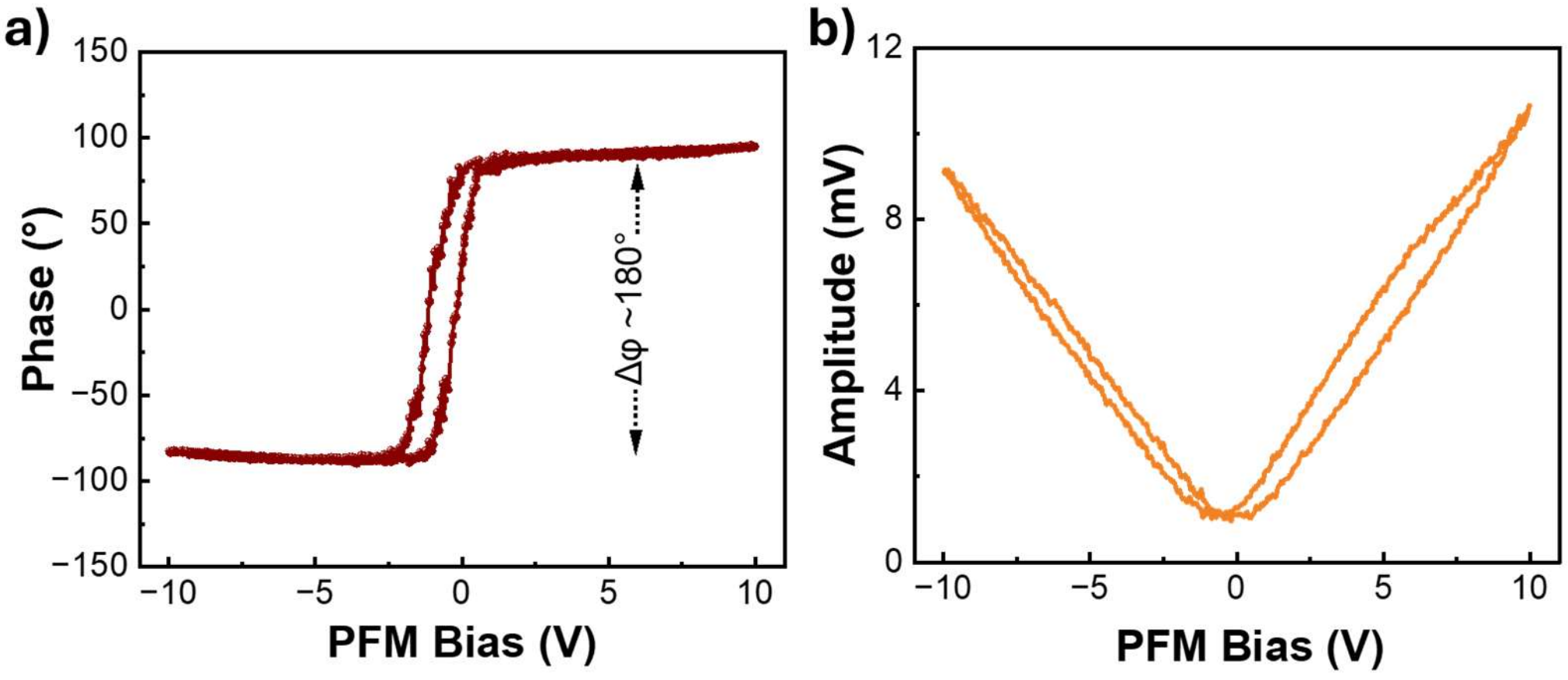

**Figure S5.** Spectroscopic LPFM switching of CA thin film as (a) ferroelectric and (b) amplitude, hysteresis.

## Additional nanogenerator device characterizations

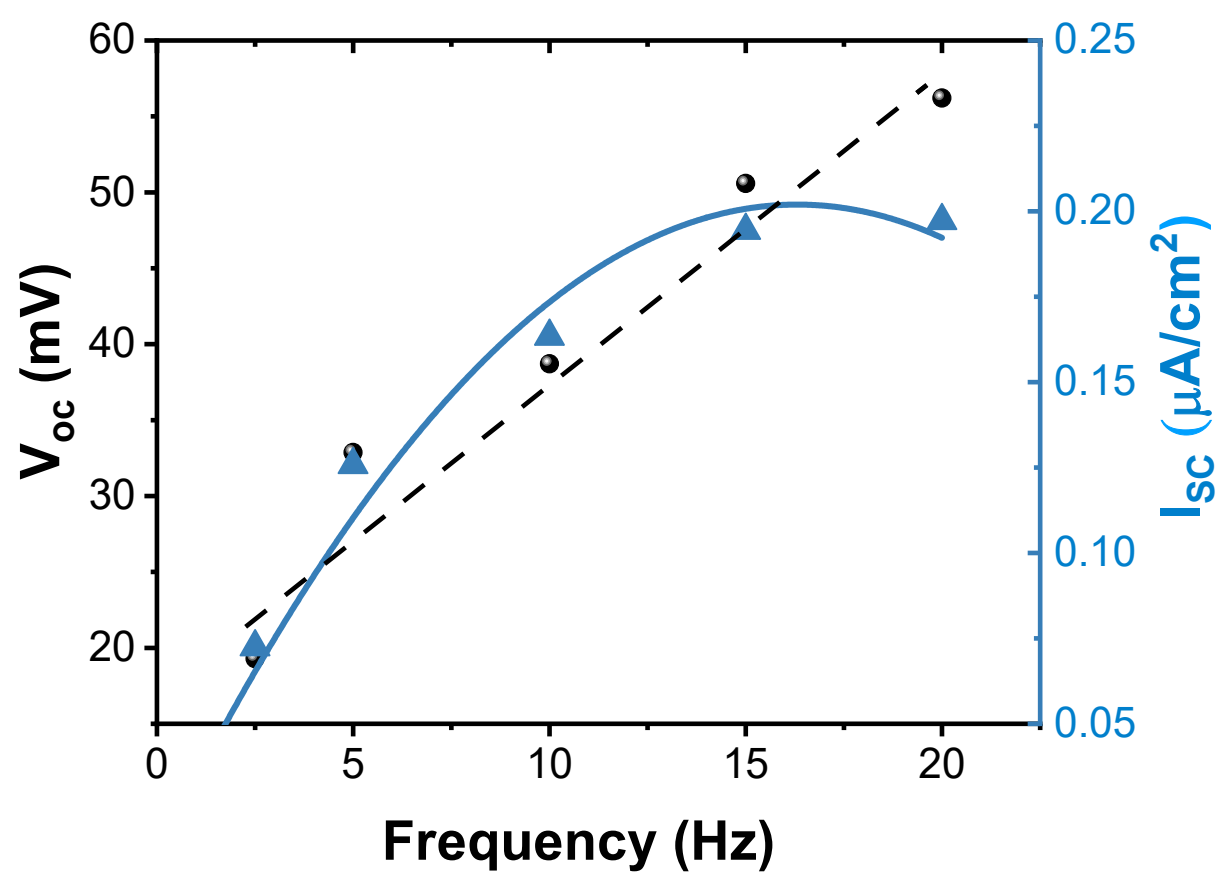

**Figure S6.** Open-circuit voltage ($V_{OC}$) and short-circuit current ($I_{SC}$) performance as a function of the cantilever vibration frequency for the CA(500 nm)-ADA(100 nm) device.

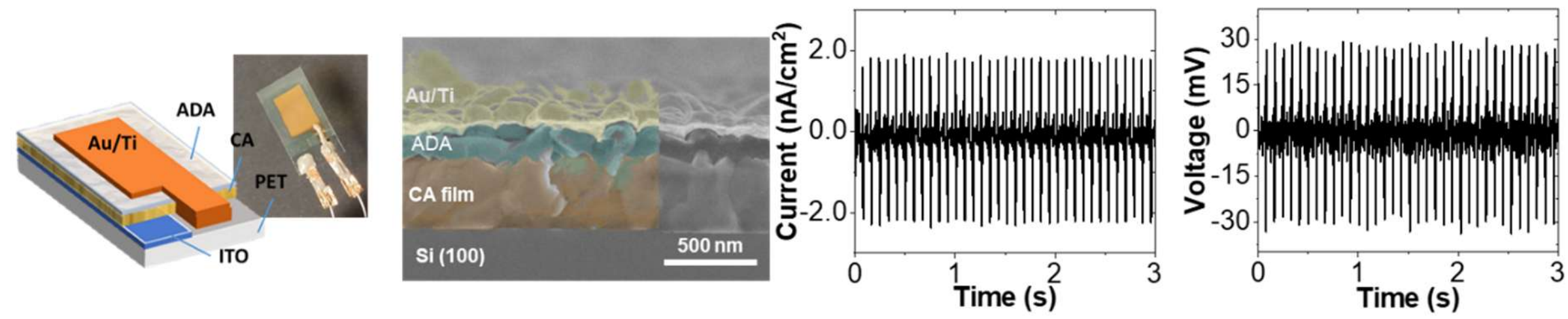

**Figure S7.** Characteristic open-circuit voltage ($V_{OC}$) (left) and short-circuit current ($I_{SC}$) (right) vs. time curves obtained for the CA (500 nm)-ADA (100 nm) piezoelectric nanogenerator operated in cantilever mode at a frequency of 11.8 Hz.

**Table S1.** Performance overview for polymeric and organic piezoelectric layers in piezoelectric nanogenerators.

| **Material systems** | **Voltage output** (V) | **Current output** (μA) | **Power output** (μW or μW/cm$^2$) | **Preparation method** | **References** |
|---|---|---|---|---|---|
| P(VDF-TrFE)<br>ZnO/P(VDF-TrFE)<br>D-phe@ZnO/P(VDF-TrFE) | 9.5<br>16.2<br>21.8 | 1.4<br>2.8<br>4.1 | 8.5 μW/cm$^2$ | Electrospnning<br>Electrospnning<br>Electrospnning | 1 |
| P(VDF-TrFE) | 5 | 1.2 | 0.8 μW/cm$^2$ | Electrospnning | 2 |
| PVDF | 14 | 29.8 | 5.1 μW/cm$^2$ | Melt-spinning PVDF fibers | 3 |
| P(VDF-TrFE) | 3 | 5.5 nA | | Template wetting | 4 |
| PVDF/Polyurethane | 3.58 | 0.286 | 0.18 μW/cm$^2$ | Electrospinning | 5 |
| PVDF–Polycarbazole | 3.8 | 2 | 0.64 μW/cm$^2$ | Electrospinning | 6 |
| Nylon-11 | 1 | 0.1 | 0.03 μW/cm$^2$ | Template-grown nanowires | 7 |
| PVDF δ-phase | 7 | 1.5 | 2 | Nanoprecipitation | 8 |
| Y-Z-PLLA | 17.52 | 2.45 | 1.76 μW/cm$^2$ | Spin-coating | 9 |
| Cellulose/PLLA | 10.3 | 0.26 | 0.4 μW | Electrospinning | 10 |
| PVDF/$SiO_2$ | 3 | 0.0185 | 10.8 μW/m$^2$ | Plasma process | 11 |
| ADA-CA | 3 | - | 30.2 μW/m$^2$ | Vacuum/Plasma process | This work |

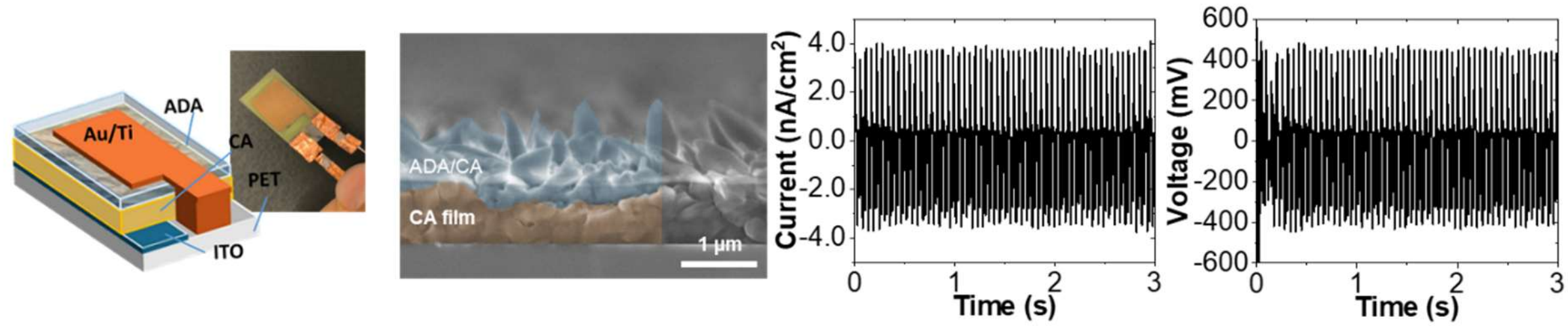

**Figure S8.** Characteristic open-circuit voltage ($V_{OC}$) (left) and short-circuit current ($I_{SC}$) (right) vs. time curves obtained for the CA (1.6 µm)-ADA (200 nm) piezoelectric nanogenerator operated in cantilever mode at a frequency of 18.3 Hz.

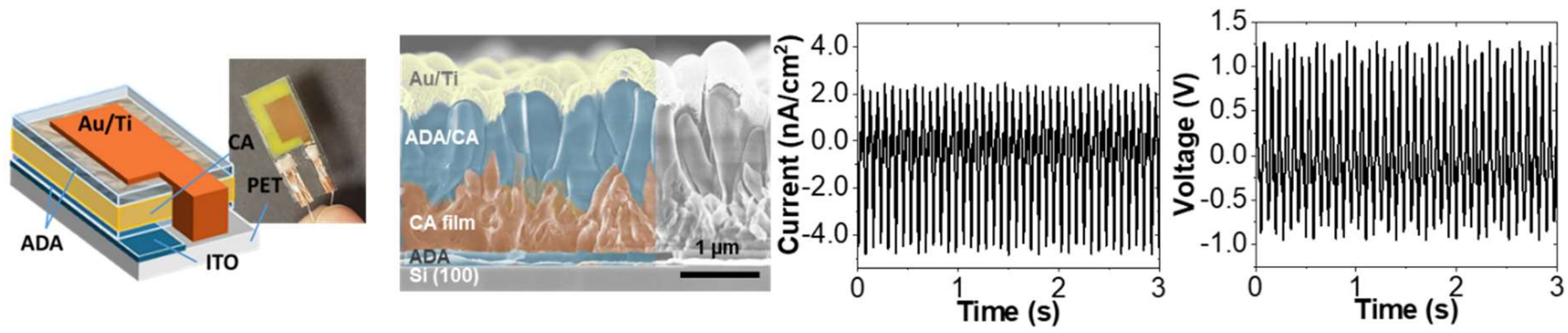

**Figure S9.** Characteristic open-circuit voltage ($V_{OC}$) (left) and short-circuit current ($I_{SC}$) (right) vs. time curves obtained for the ADA (50 nm)-CA (2 µm)-ADA (1 µm) piezoelectric nanogenerator operated in cantilever mode at a frequency of 13.3 Hz.

**Table 1 Refs.**